\documentclass[12pt]{article}
\usepackage[numbers]{natbib}
\usepackage{a4}
\usepackage{amsmath}
\usepackage{amssymb}
\usepackage{latexsym}
\usepackage{amssymb}
\usepackage{epsfig}
\usepackage{natbib}
\def \ii{{\mathrm{i}}}

\def \TP{{\mathrm{P}}}

\def \R{{\mathbb{R}}}

\def \pd{\partial}

\def \e{{\mathrm{e}}}

\def \Btau{{\boldsymbol{\tau}}}
\def \Bkappa{\boldsymbol{\kappa}}
\def \Ba{{\boldsymbol{a}}}
\def \Bb{{\boldsymbol{b}}}

\def \Bn{{\boldsymbol{n}}}

\def \Bx{{\boldsymbol{x}}}
\def \Bk{{\boldsymbol{k}}}

\def \BR{{\boldsymbol{R}}}

\begin{document}
\title{{\bf Fundamentals in generalized elasticity and dislocation theory of quasicrystals:\\
Green tensor, dislocation key-formulas and dislocation loops}}
\author{
Markus Lazar~$^\text{}$\footnote{Corresponding author.
{\it E-mail address:} lazar@fkp.tu-darmstadt.de (M.~Lazar).}
\ and
Eleni Agiasofitou~$^\text{}$\footnote{{\it E-mail address:}
agiasofitou@mechanik.tu-darmstadt.de
(E. Agiasofitou).}
\\ \\
${}^\text{}$
        Heisenberg Research Group,\\
        Department of Physics,\\
        Darmstadt University of Technology,\\
        Hochschulstr. 6,\\
        D-64289 Darmstadt, Germany\\
}

\date{\today}
\maketitle

\begin{abstract}
The present work provides fundamental quantities in generalized
elasticity and dislocation theory of quasicrystals. In a  clear and straightforward manner,
the three-dimensional Green tensor of generalized elasticity theory and
the extended displacement vector for an arbitrary extended force are
derived. Next, in the framework of dislocation theory of quasicrystals, the
solutions of the field equations for the extended displacement vector and the
extended elastic distortion tensor are given; that is the generalized Burgers
equation for arbitrary sources and the generalized Mura-Willis formula,
respectively. Moreover, important quantities of the theory of dislocations as
the Eshelby stress tensor, Peach-Koehler force, stress function tensor and the interaction energy are derived for general dislocations. The application to dislocation loops gives rise to the generalized Burgers equation, where the displacement vector can be written as a sum of a line integral plus a purely geometric part. Finally, using the Green tensor, all other dislocation key-formulas for loops, known from the theory of anisotropic elasticity,
like the Peach-Koehler stress formula, Mura-Willis equation, Volterra equation, stress function tensor and
the interaction energy are derived for quasicrystals.\\

\end{abstract}

\section{Introduction}

The knowledge of Green functions is of fundamental importance
for many physical, mathematical and engineering problems.
In the theory of partial differential equations,
a Green function is the fundamental solution of a linear
partial differential equation~(see, e.g.,~\citep{Wl}).
Using the elastic Green tensor function, one can immediately calculate the displacement field caused by
external forces in an infinite linear elastic medium. \citet{Kelvin} found the three-dimensional solution for an isotropic elastic medium.
\citet{LR} and~\citet{Synge} (see also~\citep{Barnett,Teodosiu})
derived the three-dimensional elastic Green tensor for arbitrary anisotropic materials.
\par
Quasicrystals were discovered by Shechtman in 1982
(see~\citet{Shechtman1984}).
Due to the discovery of quasicrystals, the International Union of
Crystallography changed the official definition of a crystal in 1992. For a clarification on the important subject of the definition of a quasicrystal we
refer to~\citet{Lifshitz2003, Lifshitz2007}. Shechtman was awarded the 2011 Nobel Prize in
Chemistry for his great discovery.
Quasicrystals are materials possessing long-range order but no translational
symmetry.
Nowadays, quasicrystals represent an interesting class of novel
materials. Their particular (physical, electronic, thermodynamical, chemical,
etc.) properties attract more and more the attention of researchers from
various fields and their application to several domains is highly
increasing. For instance, \citet{Kenzari2012} show that the use of quasicrystals in additive manufacturing technology has advantages compared to other composites used today, due to
their reduced friction and improved wear resistance, offering an improved functional
performance. Moreover, they show that the functional
parts contain almost no porocity and are leak-tight allowing their direct use in many fluidic applications. A systematic and comprehensive overview of the field of quasicrystals covering various aspects of the theory of elasticity and defects (cracks, dislocations) is given by~\citet{Fan11}.
\par
Three-dimensional Green functions play an important role in the theory of elasticity and defects.
They have not only pure mathematical merits themselves,
but are also important in the performance of approximative methods (finite
element method, boundary element method) as well as
in the study of cracks, dislocations and inclusions. In the literature, only some special cases
of Green functions are known for quasicrystals so far.
\citet{De} found the two-dimensional Green functions for pentagonal (2D)
quasicrystals and \citet{Ding95} calculated the explicit expressions of two-dimensional Green tensors for various forms of planar (2D) quasicrystals. \citet{Trebin98} gave an approximative solution for the three-dimensional Green tensor
 of icosahedral (3D) quasicrystals, assuming that the coupling between phonons
and phasons is small (perturbation method).
\par
In the present work, we start by deriving an analytical expression for {\it the three-dimensional elastic Green tensor for one-, two-, and three-dimensional quasicrystals}
in analogy to the theory of anisotropic elasticity, using Fourier transform. Based on the three-dimensional Green tensor another important quantity, {\it the tensor of the
potential of the second gradient of the Green tensor}, is also introduced for quasicrystals. The extended displacement vector  for an arbitrary extended force in elasticity theory of quasicrystals is also given. The mathematical structure of the higher dimensionality of quasicrystals leads in a natural way to the introduction of {\it the hyperspace notation}, which unifies the phonon and phason fields to the corresponding {\it extended field} in the hyperspace. Throughout the paper the hyperspace notation is used providing straightforward calculations.
\par
The main part of this work is devoted to the study of dislocations in quasicrystals. We generalize all the key-formulas of dislocations
known from the theory of anisotropic elasticity (e.g.,~\citep{Bacon,Balluffi,LK13})
 towards quasicrystals. In particular, we deduce the generalized
Burgers formula, Mura-Willis formula,
Peach-Koehler stress formula, Peach-Koehler force,
Eshelby stress tensor
and the interaction energy for general dislocations, that means for discrete
dislocations or a continuous distribution of dislocations.
The generalized Burgers formula
is derived following a straightforward method introduced by~\citet{LK13}
which gives directly the decomposition of the displacement vector into a part
depending on the solid angle, and a line integral part depending on the material constants.
In addition, special focus is given
on the derivation of the corresponding key-formulas for dislocation
loops. Up to now, only solutions of straight dislocations  have been found
for quasicrystals (see, e.g.,~\citep{Fan11}). It should be emphasized that the extended elastic
distortion and stress tensors  as well as the extended displacement vector produced by a dislocation loop can be written in terms of derivatives of the
three-dimensional Green tensor. In this way, the obtained dislocation key-equations build the basis of
a field theory formulation of dislocations in quasicrystals.
\par
The paper is organized as follows.
In Section 2, the basic framework of the
generalized elasticity theory of quasicrystals
with emphasis to the introduction of the hyperspace notation is presented. The
three-dimensional elastic Green tensor and the extended displacement vector
for an arbitrary extended force are derived. Section 3 is devoted to the
dislocation theory of quasicrystals. In subsections 3.1-3.5,
we derive all the dislocation key-formulas, including the $J$-integral.
Finally, subsection 3.6 provides the application to dislocation loops with the
generalized Burgers equation and all other dislocation key-formulas for
loops. Conclusions are given in Section 4. In the Appendices A and B, we give
some details about the calculation of the three-dimensional Green tensor and
its gradient.

\section{Generalized elasticity theory of quasicrystals}
\subsection{Basic framework}
This subsection is devoted to the basic framework of the generalized elasticity theory
of quasicrystals with a special focus to the introduction of the hyperspace notation. It is a compact notation which facilitates significantly the calculations throughout the paper.
\par
An $(n-3)$-dimensional quasicrystal can be generated by the projection of an
$n$-dimensional periodic structure to
the 3-dimensional physical space ($n=4,5,6$). The $n$-dimensional hyperspace $E^{n}$
can be decomposed into the direct sum of two orthogonal subspaces,
\begin{align}
\label{E-deco}
E^{n}=E_\|^3\oplus E_\perp^{(n-3)}\,,
\end{align}
where $E_\|^3$ is the 3-dimensional physical or parallel space of the
phonon fields and
$E_\perp^{(n-3)}$ is the $(n-3)$-dimensional perpendicular space of the phason
fields. For $n=4,5,6$ we speak of 1D, 2D, 3D quasicrystals and the
dimension of the hyperspace is 4D, 5D, 6D, respectively. Throughout the text,
phonon fields will be denoted by $(\cdot)^\|$ and phason fields by
$(\cdot)^\perp$.
It is important to note that all quantities (phonon and phason fields) depend on
the so-called material space coordinates $\Bx \in \R^3$.

In the theory of quasicrystals, the equilibrium conditions are of the form
(see, e.g.,~\citep{Ding1993,Hu2000})
\begin{align}
\label{EC1}
&\sigma^{\|}_{ij,j}+f_i^{\|}=0\,,\\
\label{EC2}
&\sigma^{\bot}_{ij,j}+f_i^{\bot}=0\,,
\end{align}
where $\sigma^\|_{ij}$ and $\sigma^\bot_{ij}$ are {\it the phonon and
phason stress tensors}, respectively, and $f_i^\|$ is {\it the conventional (phonon) body force density} and
$f_i^\bot$ is {\it a generalized (phason) body force density}. The comma denotes
differentiation with respect to the material coordinates. We note that the phonon stress tensor is symmetric,
$\sigma^\|_{ij}=\sigma^\|_{ji}$, while the phason stress tensor
is asymmetric, $\sigma^\bot_{ij}\neq\sigma^\bot_{ji}$ (see, e.g.,
\citep{Ding1993}).
\par
In the theory of compatible elasticity, {\it the
phonon and  phason distortion tensors}, $\beta^\|_{kl}$ and $\beta^\bot_{kl}$, are defined as the spatial
gradients of {\it the phonon and phason displacement vectors}, $u_{k}^\|$ and
$u_{k}^\bot$, respectively
\begin{align}
\label{B-u}
\beta^\|_{kl}=u_{k,l}^\|\,,\qquad
\beta^\bot_{kl}=u_{k,l}^\bot\,.
\end{align}
The constitutive relations between the stresses and distortions are
\begin{align}
\label{CR1}
&\sigma^\|_{ij}=C_{ijkl}\beta^\|_{kl}+D_{ijkl}\beta^\bot_{kl}\,,\\
\label{CR2}
&\sigma^\bot_{ij}=D_{klij}\beta^\|_{kl}+E_{ijkl}\beta^\bot_{kl}\,,
\end{align}
where $C_{ijkl}$ is the tensor of the elastic moduli of phonons,
$E_{ijkl}$ is the tensor of the elastic moduli of phasons,
and $D_{ijkl}$ is the tensor of the elastic moduli
of the phonon-phason coupling.
The constitutive tensors possess the symmetries~\citep{Ding1993}
\begin{align}
\label{C-Sym}
C_{ijkl}=C_{klij}=C_{ijlk}=C_{jikl}\,,\quad D_{ijkl}=D_{jikl}\,,
\quad E_{ijkl}=E_{klij}\,.
\end{align}
The symmetries of the tensors of the elastic constants can be simplified
  according to the specific type of the considered quasicrystal (see
  e.g. \citep{Hu2000, Fan11}).
From Eq.~(\ref{CR2}) it is obvious that the phason stress tensor
  $\sigma^\bot_{ij}$ and phason distortion tensor $\beta^\bot_{kl}$ are
  asymmetric tensors and we cannot interchange the indices $i$ with $j$ and
  $k$ with $l$, since
  the indices $i$ and $k$ ``live'' in the perpendicular space and $j$ and $l$ ``live'' in
  the material space. In general, if such indices interchange, one gets a symmetry which is sometimes called in physics a ``bastard symmetry''~\citep{GS,Hehl},
  because it interrelates two indices of totally different
  origin; for quasicrystals, namely a ``phason'' index and a
  material space index. However, such a ``bastard symmetry'' is not allowed in the theory
  of quasicrystals as it can be seen from the symmetries of the tensor
  $E_{ijkl}$ in Eq.~(\ref{C-Sym}).

If we substitute Eqs.~(\ref{CR1}), (\ref{CR2}) and (\ref{B-u}) into
Eqs.~(\ref{EC1}) and (\ref{EC2}), we obtain
{\it the coupled inhomogeneous Navier equations for the displacement vectors}
\begin{align}
\label{EOM1}
&C_{ijkl} u_{k,lj}^\|+D_{ijkl} u_{k,lj}^\bot=-f_i^\|\,,\\
\label{EOM2}
&D_{klij} u_{k,lj}^\|+ E_{ijkl} u_{k,lj}^\bot=-f_i^\bot\,.
\end{align}
\par
In what follows  we introduce {\it the hyperspace notation} for quasicrystals,
which is a compact notation in order to describe the fields in the
hyperspace.
Originally, a compact notation for the mathematical description of coupled fields was introduced by~\citet{BL75} for anisotropic linear
piezoelectric crystals. Later, this notation was generalized towards piezoelectric, piezomagnetic and magnetoelectric materials by~\citet{ADL92}.
Here, we generalize such a notation towards quasicrystals, so that
the phonon and phason fields can be unified in the corresponding extended field in the hyperspace. The components of the extended fields will be
denoted by capital letters e.g. $I,K=1,\dots,n$. Therefore, in the hyperspace
we have {\it the extended displacement vector}
\begin{align}
\label{U}
U_K=\left\{
\begin{array}{ll}
\displaystyle{u_k^{\|}}\,,\quad & \displaystyle{K=1,2,3}\,,\\
\displaystyle{u_k^{\bot}}\,,\quad &\displaystyle{K=4,\dots,n}\,, \\
\end{array}
\right.
\end{align}
{\it the extended elastic distortion tensor}
\begin{align}
\label{B}
B_{Kl}=\left\{
\begin{array}{ll}
\displaystyle{\beta_{kl}^{\|}}\,,\quad & \displaystyle{K=1,2,3}\,,\\
\displaystyle{\beta_{kl}^{\bot}}\,,\quad &\displaystyle{K=4,\dots,n}\,, \\
\end{array}
\right.
\end{align}
{\it the extended stress tensor}
\begin{align}
\label{Sigma}
\Sigma_{Ij}=\left\{
\begin{array}{ll}
\displaystyle{\sigma_{ij}^{\|}}\,,\quad & \displaystyle{I=1,2,3}\,,\\
\displaystyle{\sigma_{ij}^{\bot}}\,,\quad &\displaystyle{I=4,\dots,n}\,, \\
\end{array}
\right.
\end{align}
{\it the extended body force vector}
\begin{align}
\label{Fo}
F_I=\left\{
\begin{array}{ll}
\displaystyle{f_i^{\|}}\,,\quad & \displaystyle{I=1,2,3}\,,\\
\displaystyle{f_i^{\bot}}\,,\quad &\displaystyle{I=4,\dots,n}\,, \\
\end{array}
\right.
\end{align}
and {\it the tensor of the extended elastic moduli}
\begin{align}
\label{C}
C_{IjKl}=\left\{
\begin{array}{lll}
\displaystyle{C_{ijkl}}\,,\quad & \displaystyle{I=1,2,3}\,;\ & \displaystyle{K=1,2,3}\,,\\
\displaystyle{D_{ijkl}}\,,\quad & \displaystyle{I=1,2,3}\,;\ & \displaystyle{K=4,\dots,n}\,,\\
\displaystyle{D_{klij}}\,,\quad & \displaystyle{I=4,\dots,n}\,;\ & \displaystyle{K=1,2,3}\,,\\
\displaystyle{E_{ijkl}}\,,\quad & \displaystyle{I=4,\dots,n}\,;\ & \displaystyle{K=4,\dots,n}\,,
\end{array}
\right.
\end{align}
where $i,j,k,l=1,2,3$.
The tensor $C_{IjKl}$ retains the symmetry
\begin{align}
C_{IjKl}=C_{KlIj}\,.
\end{align}
Strictly speaking, the extended tensors appearing in Eqs.~(\ref{B}), (\ref{Sigma}) and
(\ref{C})
are called double tensor fields~\citep{Ericksen} or two-point
tensors~\citep{Marsden},
since they have indices in the hyperspace and
in the material space. In the linear theory of quasicrystals the material
space coincides with the parallel space.

In addition, in the hyperspace notation the constitutive relations~(\ref{CR1}) and (\ref{CR2}) read
\begin{align}
\label{CR-h}
\Sigma_{Ij}=C_{IjKl}\, B_{Kl}
\end{align}
and  the equilibrium conditions~(\ref{EC1}) and (\ref{EC2}) are given by
\begin{align}
\label{EC}
\Sigma_{Ij,j}+F_I=0\,.
\end{align}
By substituting Eq.~(\ref{CR-h}) into Eq.~(\ref{EC}),
the equilibrium condition reads in terms of the extended displacement vector $U_K$
\begin{align}
\label{EOM}
C_{IjKl}\, U_{K,lj}+F_I=0\,.
\end{align}
This is a Navier-type partial differential equation for the extended
displacement vector $U_K$.

\subsection{The generalized three-dimensional elastic Green tensor}
In this subsection, we derive the three-dimensional Green tensor and the
extended displacement field for an arbitrary external force for quasicrystals
in the framework of generalized elasticity theory.
\par
The method of Green functions (see, e.g.,~\citep{Wl})
is commonly used to solve linear inhomogeneous
partial differential equations like Eq.~(\ref{EOM}).
The {\it Green tensor} $G_{KM}(\BR)$ of the three-dimensional Navier equation~(\ref{EOM})
is defined by
\begin{align}
\label{GT-pde}
C_{IjKl}\, G_{KM,lj}(\BR)+\delta_{IM}\,\delta(\BR)=0\,,
\end{align}
where $\BR=\Bx-\Bx'$
and $\delta(\BR)$ is the three-dimensional Dirac delta function.
$G_{KM}(\BR)$ represents the displacement in the hyperspace
in $K$-direction at the point $\BR$
arising from a unit point force in the $M$-direction
applied at the point $\Bx'$.
The Green tensor $G_{KM}(\BR)$ satisfies the symmetry relations
\begin{align}
\label{G-sym}
G_{KM}(\BR)=G_{MK}(\BR)=G_{KM}(-\BR)\,.
\end{align}
Using the three-dimensional Fourier transform of the Green tensor
\begin{align}
\label{G-FT}
G_{KM}(\BR)=\frac{1}{(2\pi)^3}\int_{-\infty}^\infty G_{KM}(\Bk)\,
\e^{\ii \Bk\cdot\BR}\,d \Bk
\end{align}
and of the Dirac delta function
\begin{align}
\delta(\BR)=\frac{1}{(2\pi)^3}\int_{-\infty}^\infty
\e^{\ii \Bk\cdot\BR}\,d \Bk\,,
\end{align}
Eq.~(\ref{GT-pde}) can be transformed to an algebraic equation in
the Fourier space
\begin{align}
\label{GT-pde-FT}
C_{IjKl}\,k_j k_l\, G_{KM}=\delta_{IM}\,.
\end{align}
If we introduce the unit vector in the Fourier space
\begin{align}
\Bkappa=\Bk/|\Bk|
\end{align}
and the symmetric {\it Christoffel stiffness tensor} in the hyperspace
\begin{align}
(\kappa C\kappa)_{IK} = \kappa_j  C_{IjKl}\kappa_l
=\begin{pmatrix}
\kappa_j  C_{ijkl}\kappa_l & \kappa_j  D_{ijkl}\kappa_l\\
\kappa_j  D_{klij}\kappa_l & \kappa_j  E_{ijkl}\kappa_l
\end{pmatrix}\, ,
\end{align}
the Green tensor in the Fourier space is written as
\begin{align}
\label{G-k}
G_{KM}(\Bk)=\frac{1}{k^{2}}\, (\kappa C\kappa)^{-1}_{KM}\, ,
\end{align}
which is a homogeneous function of $\Bk$ of degree $-2$. The matrix $(\kappa C\kappa)_{KM}^{-1}$
is the inverse of $(\kappa C\kappa)_{KM}$
and is given by
\begin{align}
\label{C-inv}
(\kappa C\kappa)^{-1}_{KM}=\frac{A_{KM}(\Bkappa)}{D(\Bkappa)}\,,
\end{align}
where $D(\Bkappa)$ and $A_{KM}(\Bkappa)$ are the determinant and the adjoint
of the matrix $(\kappa C\kappa)_{KM}$, respectively.
Substituting Eq.~(\ref{G-k}) into Eq.~(\ref{G-FT}),
the three-dimensional Fourier integral can be reduced to a line integral along the unit circle in the plane orthogonal to $\BR$
(see Appendix~\ref{appendixA} and \citep{Synge,Teodosiu,Balluffi})
\begin{align}
\label{G-int}
\int_{-\infty}^\infty G_{KM}(\Bk)\,
\e^{\ii \Bk\cdot\BR}\,d \Bk=\frac{\pi}{R}\int_0^{2\pi} G_{KM}(\Bn)\,d \phi\,,
\end{align}
where $\Bn$ is a unit vector ``scanning'' the circle of integration and
remaining orthogonal to $\BR$, thus $\Bn\cdot \BR=0$.
In this way, the Green tensor~(\ref{G-FT}) can be written in the form
(see Appendix~\ref{appendixA})
\begin{align}
\label{GT}
G_{KM}(\BR)=
\frac{1}{8\pi^2 R}\, \int_0^{2\pi} (n C n)_{KM}^{-1}\, d \phi\, .
\end{align}
Here, $\Bn$ is a function of $\phi$.
Eq.~(\ref{GT}) is {\it the three-dimensional elastic Green tensor
for quasicrystals}.
The Green tensor~(\ref{GT}) is the generalization of the Green tensor
of general anisotropic elasticity
(see, e.g.,~\citep{Synge,Barnett,Teodosiu,LK13,Balluffi,Li})
towards quasicrystals.
The integral in Eq.~(\ref{GT}) can be computed by standard numerical methods, when
$C_{IjKm}$ is given (see, e.g., \citep{Barnett,Bacon})
 and therefore it is well suited to rapid and accurate numerical
integration.
The numerical calculation of the Green tensor function of an infinite
quasicrystalline medium with general anisotropy can be reduced to the
application of standard numerical codes. The elastic Green tensor $G_{KM}(\BR)$ can be decomposed into its phonon and
phason parts
\begin{align}
G_{KM}(\BR)
=\begin{pmatrix}
G^{\| \|}_{km}(\BR)&
G^{\|\perp}_{km}(\BR)\\
G^{\perp\|}_{km}(\BR)&
G^{\perp\perp}_{km}(\BR)\\
\end{pmatrix}
\end{align}
and has the following physical interpretations:
\begin{align}
G^{\| \|}_{km}(\BR)=\ & \text{phonon displacement at $\Bx$ in the direction $x_k$
due to a unit}
\nonumber\\
& \text{phonon point force at $\Bx'$ in the $x_m$ direction;}
\nonumber\\
G^{\| \perp}_{km}(\BR)=\ & \text{phonon displacement at $\Bx$ in the direction $x_k$
due to a unit}
\nonumber\\
& \text{phason point force at $\Bx'$ in the $x_m$ direction;}
\nonumber\\
G^{\perp \|}_{km}(\BR)=\ & \text{phason displacement at $\Bx$ in the direction $x_k$
due to a unit}
\nonumber\\
& \text{phonon point force at $\Bx'$ in the $x_m$ direction;}
\nonumber\\
G^{\perp \perp}_{km}(\BR)=\ & \text{phason displacement at $\Bx$ in the direction $x_k$
due to a unit}
\nonumber\\
& \text{phason point force at $\Bx'$ in the $x_m$ direction.}
\nonumber
\end{align}

The solution of the Green tensor
in quasicrystalline materials can be applied to calculate phonon and
phason fields caused by an external or internal force.
For an arbitrary extended force $F_M$, the particular solution of
Eq.~(\ref{EOM}) is written as
\begin{align}
\label{U-sol-F}
U_K(\Bx)=G_{KM}*F_M,
\end{align}
where the symbol $*$ denotes the three-dimensional spatial convolution. Using the Green tensor~(\ref{GT}), Eq.~(\ref{U-sol-F}) gives {\it the extended displacement vector for an arbitrary extended force $F_M$ in elasticity theory of quasicrystals}
\begin{align}
\label{U-sol-F1}
U_K(\Bx)=\frac{1}{8\pi^2}\,\int_V \frac{F_M(\Bx')}{R} \bigg(\int_0^{2\pi} (n C n)_{KM}^{-1}\, d \phi\bigg)\, d V'\,.
\end{align}
For instance, for a Kelvin-type force, that is $F_M(\BR)=f_M \delta(\BR)$ with constant magnitude $f_M$, Eq.~(\ref{U-sol-F1}) reduces to
\begin{align}
\label{U-sol-F2}
U_K(\Bx)=G_{KM}(\Bx)f_M=\frac{f_M}{8\pi^2 r}\,\int_0^{2\pi} (n C n)_{KM}^{-1}\, d \phi\,.
\end{align}
 Eq.~(\ref{U-sol-F2}) is {\it the Kelvin-type force solution for quasicrystals}.

\section{Dislocation theory of quasicrystals}
\subsection{Basic framework, the field equations and their solutions}
First, the basic framework for dislocations in quasicrystals is
presented. Next, we give the field equations for the extended displacement
vector and the extended elastic distortion tensor and we derive their
particular solutions.
For a review of the physics of dislocations in
quasicrystals, we refer to~\citet{Feuerb12} and \citet{WH}.
\par
In general, if dislocations are present, the theory of compatible elasticity modifies to the theory
of incompatible elasticity incorporating plastic fields.
For incompatible elasticity theory of quasicrystals we refer
to~\citet{Ding95, Hu2000} and \citet{ALK2010}.
In the presence of dislocations inside the medium, the displacement gradient
is usually decomposed into {\it the elastic distortion tensors} $\beta^\|_{ij}$, $\beta^\bot_{ij}$,
and {\it the plastic distortion tensors} ${\beta}^{\|\, \TP}_{ij}, \ {\beta}^{\bot\, \TP}_{ij}$,
according to
\begin{align}
\label{u-grad}
u^\|_{i,j}=\beta^\|_{ij}+{\beta}^{\|\, \TP}_{ij}\,, \qquad
u^\bot_{i,j}=\beta^\bot_{ij}+{\beta}^{\bot\, \TP}_{ij}\, .
\end{align}
The incompatibility of the elastic and plastic parts gives rise to the existence of dislocation density tensors.{\it The phonon and phason dislocation density tensors} $\alpha^\|_{ij}$ and $\alpha^\bot_{ij}$, respectively, are defined in terms of the elastic distortion tensors
\begin{align}
&\alpha^\|_{ij}=\epsilon_{jkl}\beta^\|_{il,k}\, ,  \qquad
\alpha^\bot_{ij}=\epsilon_{jkl}\beta^\bot_{il,k}
\label{DD}
\end{align}
or in terms of the plastic distortion tensors
\begin{align}
\alpha^\|_{ij}=-\epsilon_{jkl}\beta^{\|\, \TP}_{il,k}\, ,
\qquad     \alpha^\bot_{ij}=-\epsilon_{jkl}\beta^{\bot\,\TP}_{il,k}\,,
\label{DD-p}
\end{align}
where $\epsilon_{jkl}$ is the three-dimensional Levi-Civita tensor. In our notation, the first index of the dislocation density tensor (Eq.~(\ref{DD-p})) shows the orientation of the Burgers vector and the second index shows the direction of the dislocation line.
The dislocation density tensors satisfy the following Bianchi identities
\begin{align}
\alpha^\|_{ij,j}=0\,, \qquad   \alpha^\bot_{ij,j}=0\,,
\label{BI}
\end{align}
which mean that dislocations cannot end inside the quasicrystalline medium.
\par
In the absence of external forces, the field equations for the phonon and phason displacement fields are (see, e.g.,~\citep{Ding95,ALK2010,AL13})
\begin{align}
\label{u-pho}
&C_{ijkl}u_{k,lj}^\|+D_{ijkl}u_{k,lj}^\bot= C_{ijkl}{\beta}^{\|\, \TP}_{kl,j}+D_{ijkl}{\beta}^{\bot\, \TP}_{kl,j}\,,\\
\label{u-pha}
&D_{klij}u_{k,lj}^\|+E_{ijkl}u_{k,lj}^\bot=D_{klij}{\beta}^{\|\, \TP}_{kl,j}+E_{ijkl}{\beta}^{\bot\, \TP}_{kl,j}\,,
\end{align}
where the plastic distortion tensors play the role of the sources for the
displacement vectors.
The corresponding field equations for the elastic distortion tensors are of the form~\citep{AL13}
\begin{align}
\label{B-pho}
&C_{ijkl}\beta_{km,lj}^\|+D_{ijkl}\beta_{km,lj}^\bot=
\epsilon_{lmp}\big(C_{ijkl}{\alpha}^\|_{kp,j}+D_{ijkl}{\alpha}^\bot_{kp,j}\big)\,,\\
\label{B-pha}
&D_{klij}\beta_{km,lj}^\|+E_{ijkl}\beta_{km,lj}^\bot=
\epsilon_{lmp}\big(D_{klij}{\alpha}^\|_{kp,j}+E_{ijkl}{\alpha}^\bot_{kp,j}\big)\,,
\end{align}
where the dislocation density tensors are the source fields.
\par
A dislocation in a quasicrystal can be considered as a ``hyperdislocation" in the hyperlattice by means of a generalized Volterra process. Because the hyperlattice is periodic, the generalized Volterra process can be understood as insertion or removal of a hyper-halfplane (e.g.,~\citep{Feuerb12}). The Burgers vector of the ``hyperdislocation" consists of phonon and phason components
\begin{equation}
\label{BV}
b_I=(b_i^\|, b_i^\bot) \ \in E_\| \oplus E_\bot\,.
\end{equation}
A ``hyperdislocation" is a line defect in a quasicrystal characterized by the Burgers vector and the direction of the dislocation line in the material space. It should be noted that
for a perfect dislocation in a quasicrystal, both
components $\Bb^\|$ and $\Bb^{\perp}$ are non-zero
and the Burgers vector $\Bb$  is a lattice vector in the hyperspace.
If the phason component $\Bb^{\perp}$ is zero, then there exist a stacking
fault along the cutting surface of the generalized Volterra process, and
this dislocation represents a partial dislocation,
since $\Bb^\|$ alone is not a lattice vector in the hyperspace (see~\citep{WH}).

Using the hyperspace notation, we can write {\it the extended plastic distortion tensor}
\begin{align}
\label{BP}
B_{Kl}^\TP=\left\{
\begin{array}{ll}
\displaystyle{\beta_{kl}^{\|\, \TP}}\,,\quad & \displaystyle{K=1,2,3}\,,\\
\displaystyle{\beta_{kl}^{\bot\, \TP}}\,,\quad &\displaystyle{K=4,\dots,n}\,, \\
\end{array}
\right.
\end{align}
and {\it the extended dislocation density tensor}
\begin{align}
\label{A}
A_{Kl}=\left\{
\begin{array}{ll}
\displaystyle{\alpha_{kl}^{\|}}\,,\quad & \displaystyle{K=1,2,3}\,,\\
\displaystyle{\alpha_{kl}^{\bot}}\,,\quad &\displaystyle{K=4,\dots,n}\,. \\
\end{array}
\right.
\end{align}
With the definitions (\ref{BP}) and (\ref{A}), Eqs.~(\ref{u-grad})--(\ref{DD-p}) can be respectively written
\begin{align}
\label{U-deco}
U_{I,j}=B_{Ij}+B^\TP_{Ij}
\end{align}
and
\begin{align}
\label{A2}
A_{Ij}=\epsilon_{jkl}B_{Il,k}\,,\qquad
A_{Ij}=-\epsilon_{jkl}B^\TP_{Il,k}\,.
\end{align}
Moreover, the Bianchi identity is reduced (in the hyperspace) to
\begin{align}
\label{BI-h}
A_{Kl,l}=0\,.
\end{align}
\par
In the hyperspace, {\it the field equation for the extended displacement vector} $U_K$ (see Eqs. (\ref{u-pho}) and (\ref{u-pha})) is written as
\begin{align}
\label{U-EOM}
C_{IjKl}\, U_{K,lj}=C_{IjKl} B^\TP_{Kl,j}\,,
\end{align}
which is an inhomogeneous Navier equation.
If we compare Eqs.~(\ref{U-EOM}) and (\ref{EOM}),
we may introduce an ``internal force caused by dislocations''
as
\begin{align}
\label{F-d}
F_I=-C_{IjKl} B^\TP_{Kl,j}\,.
\end{align}
The force~(\ref{F-d}) is a fictitious body force.
From Eq.~(\ref{F-d}), the internal phason force density due to dislocations
reads (see also~\citep{Ding95, Hu2000})
\begin{align}
\label{F-phason}
f^\bot_i=-D_{klij} \beta^{\|\, \TP}_{kl,j}-E_{ijkl} \beta^{\bot\, \TP}_{kl,j}
\,,
\end{align}
which is non-zero if dislocations exist in the quasicrystalline medium.
Thus, Eq.~(\ref{F-phason}) is an example of a phason force
caused by the gradient of the plastic fields of dislocations.
The particular solution of Eq.~(\ref{U-EOM}) following Eq.~(\ref{U-sol-F}) reads
\begin{align}
\label{U-sol}
U_{I}=G_{IJ}*F_J=-C_{JkLm} G_{IJ} * B^\TP_{Lm,k}=-C_{JkLm} G_{IJ,k} * B^\TP_{Lm}\,,
\end{align}
where $G_{IJ}$ is given by Eq.~(\ref{GT}).
Eq.~(\ref{U-sol}) gives {\it the extended displacement vector in a quasicrystal that has experienced
an extended plastic distortion $B^\TP_{Lm}$.}
It is important to note that Eq.~(\ref{U-sol}) is the Volterra-type
representation of the displacement vector $U_I$ for an arbitrary plastic
distortion  tensor $B^\TP_{Lm}$.

\par
The field equations~(\ref{B-pho}) and (\ref{B-pha}) simplify in the hyperspace notation to the following {\it field equation for the extended elastic distortion tensor}
\begin{align}
\label{B-EOM}
C_{IjKl} B_{Km,lj}=\epsilon_{lmn}\, C_{IjKl} A_{Kn,j}\,,
\end{align}
which is a tensorial Navier equation.
The particular solution of Eq.~(\ref{B-EOM}) is given by
\begin{align}
\label{B-sol}
B_{Im}=\epsilon_{mnr}C_{JkLn} G_{IJ,k} *A_{Lr}\,.
\end{align}
Eq.~(\ref{B-sol}) is  {\it the generalization of the so-called Mura-Willis
  formula}~\citep{deWit2, Willis67} {\it towards quasicrystals.} Once we know the extended elastic distortion tensor we can calculate the extended stress tensor by means of Eq.~(\ref{CR-h}), that is
\begin{align}
\label{T-sol}
\Sigma_{Ps}=C_{PsIm}\epsilon_{mnr}C_{JkLn} G_{IJ,k} *A_{Lr}\,.
\end{align}

\subsection{The generalized Burgers equation for arbitrary sources}
 In this subsection, we derive an alternative expression for the solution
of the extended displacement vector $U_I$, using a straightforward method introduced by~\citet{LK13}.
This method gives directly
the Burgers equation for arbitrary plastic distortions and dislocation
densities. It is mainly based on three steps: the linear decomposition of the total distortion
into the elastic and plastic distortions, the Green function of the
Poisson equation and the Mura-Willis formula. Using this method, the solution of the extended displacement field can be decomposed into a purely geometric part depending on the
plastic distortion and a part depending on the tensor of the elastic constants
and the dislocation density. Therefore, it is not necessary to solve the
inhomogeneous Navier equation~(\ref{U-EOM})
in order to extract a purely geometric part from it.
\par
The divergence from the right of Eq.~(\ref{U-deco}) gives the following
Poisson equation for the displacement field $U_I$
\begin{align}
\label{U-Laplace}
\Delta U_I=B^\TP_{Im,m}+B_{Im,m}\,.
\end{align}
Using the three-dimensional Green function of the Poisson equation
(e.g.,~\citep{Wl, Barton})
\begin{align}
\label{GF-laplace}
\Delta G=\delta(\BR)\,,\qquad
G=-\frac{1}{4\pi R}\,,
\end{align}
the solution $U_I$ of Eq.~(\ref{U-Laplace}) is given by
\begin{align}
\label{U-sol1}
U_I=-\big[B^\TP_{Im,m}+B_{Im,m}\big]*\frac{1}{4\pi R}\,.
\end{align}
The above equation using the generalized Mura-Willis formula~(\ref{B-sol}) is written
\begin{align}
\label{U-sol2}
U_I=-B^\TP_{Im,m}*\frac{1}{4\pi R}
-\big[\epsilon_{mnr}C_{JkLn} G_{IJ,km} *A_{Lr}\big]*\frac{1}{4\pi R}\,.
\end{align}
Using the associative law for the convolution,
Eq.~(\ref{U-sol2}) can be rewritten as
\begin{align}
\label{U-sol3}
U_I=-B^\TP_{Im,m}*\frac{1}{4\pi R}
-\epsilon_{mnr}C_{JkLn}\Big[G_{IJ,km} *\frac{1}{4\pi R}\Big]*A_{Lr}\,.
\end{align}
The inconvenience of the double convolution in the second term of Eq.~(\ref{U-sol3})
can be reduced to a single one.
To this aim, we introduce the tensor $F_{mnIJ}$, which was originally
introduced by~\citet{Kirchner,Kirchner2} for anisotropic elasticity (see also \citep{LK13})
\begin{align}
\label{F}
F_{mkIJ}=G_{IJ,km}*\frac{1}{4\pi R}\,.
\end{align}
We may call the tensor $F_{mnIJ}$ as {\it the potential of the second gradient
of the Green tensor}, since it satisfies the Poisson equation
\begin{align}
\Delta F_{mkIJ}+G_{IJ,km}=0\,.
\end{align}
Moreover, the following relationships hold
\begin{align}
F_{mkIJ,m}+G_{IJ,k}=0\,,\qquad
F_{mkIJ,k}+G_{IJ,m}=0\,
\end{align}
and
\begin{align}
F_{mkIJ,mk}+\Delta G_{IJ}=0\,.
\end{align}
In addition, $F_{mnIJ}$ possesses the symmetry properties
\begin{align}
\label{F-sym}
F_{mkIJ}=F_{kmIJ}=F_{mkJI}\,
\end{align}
and
\begin{align}
F_{mkIJ}(\BR)=F_{mkIJ}(-\BR)\,.
\end{align}
Using Eq.~(\ref{GT-pde}), Eq.~(\ref{F}) becomes
\begin{align}
\label{F-C}
C_{JkLm}F_{mkIJ}=-\frac{1}{4\pi R}\,\delta_{IL}\,.
\end{align}
The Fourier transform  of $F_{mnIJ}$ is
\begin{align}
\label{F-k}
F_{mkIJ}(\Bk)=-\frac{1}{k^{2}}\, \kappa_m\kappa_k (\kappa C\kappa)^{-1}_{IJ}\, .
\end{align}
Like the Fourier transform of the Green tensor~(\ref{G-k}), $F_{mnIJ}$ in
Eq.~(\ref{F-k}) varies like $k^{-2}$.
Thus, its three-dimensional inverse Fourier transform for $\Bn\cdot\BR=0$
reduces to a one-dimensional integration in the angle $\phi$
\begin{align}
\label{F-int}
\int_{-\infty}^\infty F_{mkIJ}(\Bk)\,
\e^{\ii \Bk\cdot\BR}\,d \Bk=\frac{\pi}{R}\int_0^{2\pi} F_{mkIJ}(\Bn)\,d \phi\,,
\end{align}
where $\Bn$ is a unit vector ``scanning'' the circle of integration and
remaining orthogonal to $\BR$.
Consequently, the three-dimensional potential of the second gradient
of the Green tensor is given by
\begin{align}
\label{F-int2}
F_{mkIJ}(\BR)=
-\frac{1}{8\pi^2 R}\, \int_0^{2\pi} n_m n_k (n C n)_{IJ}^{-1}\, d \phi\, .
\end{align}
Like Eq.~(\ref{GT}), Eq.~(\ref{F-int2})
is well suited to rapid and accurate numerical integration. Moreover, comparing Eqs.~(\ref{GT}) and (\ref{F-int2}), we obtain the following important relation
\begin{align}
\label{F-tr}
\delta_{mk} F_{mkIJ}=-G_{IJ}\,.
\end{align}
Hence, Eq.~(\ref{U-sol3}) via the definition ~(\ref{F}) reduces to a single convolution integral
\begin{align}
\label{U-final}
U_I=-B^\TP_{Im,m}*\frac{1}{4\pi R}-\epsilon_{rmn} C_{JkLn} F_{mkIJ}* A_{Lr}\,,
\end{align}
where $F_{mkIJ}$ is given by Eq.~(\ref{F-int2}). Eq.~(\ref{U-final}) is {\it the generalized Burgers equation for arbitrary sources.} It represents {\it the solution of the extended displacement vector in dislocation theory}, when $B^\TP_{IJ}$ and $A_{Ij}$ are
given and is valid for any distribution of dislocations. The plastic distortion tensor
$B^\TP_{IJ}$ and the corresponding dislocation density tensor $A_{Ij}$ can represent
discrete dislocations (straight dislocations, dislocation loops) or a
continuous distribution of dislocations. It is worth noting that the first
term in Eq.~(\ref{U-final}) is a purely geometric part because it does not
depend on the properties of the medium. Only the second part depends on the
properties of the material due to the appearance of the tensor of elastic
constants $C_{JkLn}$ and the tensor $F_{mkIJ}$.
The structure of Eq.~(\ref{U-final}) is a direct consequence of the
decomposition of the total distortion tensor into an elastic and a plastic
part~(see Eq.~(\ref{U-deco})).
Due to this clear decomposition of the displacement vector
into a geometric part determined by the plastic distortion and a part
depending on the elastic coefficients and the dislocation density tensor,
the representation of the displacement field ~(\ref{U-final}) is more suitable than the Volterra-type representation~(\ref{U-sol})
in solving dislocation problems.

\subsection{The Eshelby stress tensor and the Peach-Koehler force}

 We derive here quantities that play an important role in defect mechanics
 and in the so-called Eshelbian mechanics \citep{Markenscoff,Maugin93,Kienzler,Maugin11};
namely the Eshelby stress tensor, the Peach-Koehler force and the $J$-integral.
In general, the $J$-integral \citep{Rice73,Eshelby75,Kirchner99} is important for
dislocations, cracks and fracture mechanics, especially for a dislocation based fracture mechanics.

We start with a direct derivation of the Eshelby stress tensor of quasicrystals
following~\citet{LK13}.
If we multiply Eq.~(\ref{A2}) by $\epsilon_{jkl}$, we obtain
\begin{align}
\label{alpha2}
B_{Ik,j}-B_{Ij,k}=\epsilon_{jkl} A_{Il}\,,
\end{align}
which  multiplied by $\Sigma_{Ik}$ gives
\begin{align}
W{,_j}-\Sigma_{Ik} B_{Ij,k}=\epsilon_{jkl}\Sigma_{Ik} A_{Il}\,,
\end{align}
where the elastic strain energy density (for the unlocked state) is given by
\begin{align}
\label{W}
W=\frac{1}{2}\,\Sigma_{Ij} B_{Ij}=
\frac{1}{2}\,\sigma^{\|}_{ij} \beta^{\|}_{ij}
+\frac{1}{2}\,\sigma^{\perp}_{ij} \beta^{\perp}_{ij}\,.
\end{align}
Using Eq.~(\ref{EC}) for vanishing extended body forces, we obtain
\begin{align}
\label{PK0}
\big[W\delta_{jk}-\Sigma_{Ik} B_{Ij}\big]{,_k}=
\epsilon_{jkl}\Sigma_{Ik}A_{Il}\,.
\end{align}
In the brackets on the left hand side
of Eq.~(\ref{PK0}), {\it the Eshelby stress tensor for quasicrystals}~\citep{ALK2010} appears
\begin{align}
\label{P}
P_{jk}=W\delta_{jk}-\Sigma_{Ik} B_{Ij}
=W\delta_{jk}-\sigma^{\|}_{lk} \beta^{\|}_{lj}
-\sigma^{\perp}_{lk} \beta^{\perp}_{lj}\,
\end{align}
and it consists of phonon and phason fields.
The trace of the Eshelby stress tensor~(\ref{P})
reads
\begin{align}
\label{P-tr}
P_{jj}=\frac{1}{2}\, \Sigma_{Ij} B_{Ij}
=\frac{1}{2}\,\big(\sigma^{\|}_{ij} \beta^{\|}_{ij}
+\sigma^{\perp}_{ij} \beta^{\perp}_{ij}\big)\,.
\end{align}
The skew-symmetric part of the  Eshelby stress tensor~(\ref{P})
is given by
\begin{align}
\label{P-sk}
\epsilon_{ijk}P_{jk}=\epsilon_{ijk}\Sigma_{Ij} B_{Ik}
=\epsilon_{ijk}\big(\sigma^{\|}_{lj} \beta^{\|}_{lk}
+\sigma^{\perp}_{lj} \beta^{\perp}_{lk}\big)\,.
\end{align}
It is known that the Eshelby stress tensor stems from  spatial translational transformations in the
material space
and is the static part of the energy-momentum tensor. The source term on the right hand side of Eq.~(\ref{PK0})
is the so-called {\it Peach-Koehler force density}~\citep{ALK2010}
\begin{align}
\label{PK}
f^{\text{PK}}_j=\epsilon_{jkl}\Sigma_{Ik} A_{Il}
=\epsilon_{jkl}\big(\sigma^{\|}_{ik} \alpha^{\|}_{il}
+\sigma^{\perp}_{ik} \alpha^{\perp}_{il}\big)\,,
\end{align}
which consists of a phonon stress-dislocation density part and a
phason stress-dislocation density part.
Thus, Eq.~(\ref{PK0}) is a translational material balance law, where the divergence of the Eshelby stress tensor~(\ref{P})
is balanced by the Peach-Koehler force density~(\ref{PK}),
\begin{align}
\label{BL}
P_{jk,k}=f^{\text{PK}}_j\,.
\end{align}
\par
The integral form of the balance law~(\ref{BL}) gives the so-called {\it $J$-integral for dislocations in  quasicrystals}
\begin{align}
\label{J}
J_j:=\int_V P_{jk,k} \, d V=\int_S P_{jk}\, d S_k
=\int_S \big[W\delta_{jk}-\Sigma_{Ik} B_{Ij}\big] \, d S_k
=\int_V f^{\text{PK}}_j\, d V\,,
\end{align}
where the Gauss theorem has been used.
From Eq.~(\ref{J}) it can be seen that the $J$-integral for dislocations is
equivalent to the Peach-Koehler force (see also \citep{Rice85,Weertman96,Kirchner99}).

\subsection{The stress function tensor}
Herein, we deduce the stress function tensor of first order for the self-stresses
of general dislocations.
Using the method of the stress function tensor of first order
(e.g.,~\citep{Kroener58}),
the equilibrium condition~(\ref{EC}) is fulfilled
automatically for vanishing external forces.
\par
For the self-stresses caused by dislocations,
that means that the external forces are zero, the equilibrium condition~(\ref{EC})
can be satisfied by deriving the asymmetric stress $\Sigma_{Ij}$
from an asymmetric {\it stress function tensor} $\Phi_{Ij}$, which is a
stress function tensor of first order, as follows
\begin{align}
\label{SF}
\Sigma_{Ij}=\epsilon_{jkl} \Phi_{Il,k}\,.
\end{align}
For the stress~(\ref{SF}), the equilibrium condition~(\ref{EC}) for vanishing
forces, $\Sigma_{Ij,j}=0$, is automatically satisfied. Now, we perform the curl on the right index of the stress tensor~(\ref{SF})
\begin{align}
\epsilon_{mnj} \Sigma_{Ij,n}
=\epsilon_{mnj}\epsilon_{jkl}\Phi_{Il,kn}
=\Phi_{Il,lm}-\Delta\Phi_{Im}\,.
\end{align}
Imposing the side condition (see, e.g.,~\citep{Kroener58})
\begin{align}
\label{phi-cond3}
\Phi_{Ij,j}=0\,,
\end{align}
we find the following Poisson equation for the stress function tensor
\begin{align}
\label{phi-laplace}
\Delta\Phi_{Ij}=-\epsilon_{jkl}\Sigma_{Il,k}\,.
\end{align}
Using Eq.~(\ref{GF-laplace}), we find the solution
\begin{align}
\label{phi-sol1}
\Phi_{Ij}=\epsilon_{jkl} \Sigma_{Il,k}*\frac{1}{4\pi R}\,,
\end{align}
which with the help of the constitutive relation~(\ref{CR-h}) becomes
\begin{align}
\label{phi-sol}
\Phi_{Ij}=\epsilon_{jkl}C_{IlMn} B_{Mn,k}*\frac{1}{4\pi R}\,.
\end{align}
Substituting Eq.~(\ref{B-sol}) into Eq.~(\ref{phi-sol}) and using
the associative law for the convolution (see, e.g.,~\citep{Wl}), we obtain
\begin{align}
\label{phi-sol2}
\Phi_{Ij}&=\Big[\epsilon_{jkl}C_{IlMn}\epsilon_{npq}C_{RsTp}\,
G_{MR,ks} *\frac{1}{4\pi R}\Big]* A_{Tq}\,.
\end{align}
It is easy to check that Eq.~(\ref{phi-sol2}) satisfies the side condition~(\ref{phi-cond3}) and reproduces the extended stress tensor~(\ref{T-sol}) by inserting Eq.~(\ref{phi-sol2}) into Eq.~(\ref{SF}). Moreover, one can see that in Eq.~(\ref{phi-sol2}) the potential of the second gradient of the Green tensor $F_{mkIJ}$ (Eq.~(\ref{F})) is appearing. Consequently,  Eq.~(\ref{phi-sol2}) is rewritten as
\begin{align}
\label{phi-sol3}
\Phi_{Ij}&=\epsilon_{jkl}C_{IlMn}\epsilon_{npq}C_{RsTp}\,
F_{skMR} *A_{Tq}\,.
\end{align}
Eq.~(\ref{phi-sol3}) gives {\it the stress function tensor for dislocations in quasicrystals} and it holds for discrete dislocations as well as for a continuous distribution of dislocations. Furthermore,  the product of the two Levi-Civita tensors may be factored out and
if we use Eqs.~(\ref{BI-h}) and (\ref{F-C}), Eq.~(\ref{phi-sol3}) simplifies to
\begin{align}
\label{phi-sol4}
\Phi_{Ij}&=
C_{IlTl}\, \frac{1}{4\pi R}* A_{Tj}
-C_{IlTj}\, \frac{1}{4\pi R}* A_{Tl}
+C_{IlMk}C_{RsTl}\, F_{skMR}* A_{Tj}
-C_{IlMk}C_{RsTj}\, F_{skMR}* A_{Tl}
\,.
\end{align}

\subsection{The interaction energy}
We calculate the interaction energy between two general dislocations, that means discrete dislocations or a continuous distribution of dislocations.
\par
The interaction energy between two dislocations is defined by
\begin{align}
\label{W-int}
W^{(AB)}=\int_V \Sigma^{(B)}_{Ij} B^{(A)}_{Ij}\, d V\,,
\end{align}
where $\Sigma^{(A)}_{Ij}$, $\Sigma^{(B)}_{Ij}$ are the (asymmetric) extended
stress tensors and $B^{(A)}_{Ij}$, $B^{(B)}_{Ij}$ are the extended elastic distortions of the individual dislocations. Expressing the stresses  in terms of the corresponding stress function tensors  $\Phi^{(A)}_{Ij}$, $\Phi^{(B)}_{Ij}$ (see Eq.~(\ref{SF})), using integration by parts and neglecting the surface terms at infinity, we obtain
\begin{align}\label{W-int1}
W^{(AB)}=\int_V \big(\epsilon_{jkl} \Phi^{(B)}_{Il,k}\big) B^{(A)}_{Ij}\, d V
=\int_V \Phi^{(B)}_{Il}\, \big(\epsilon_{lkj} B^{(A)}_{Ij,k}\big)\, d V
=\int_V \Phi^{(B)}_{Ij} A^{(A)}_{Ij}\, d V\,,
\end{align}
where $A^{(A)}_{Ij}$, $A^{(B)}_{Ij}$ are the corresponding extended dislocation densities. If we substitute Eq.~(\ref{phi-sol3}) into Eq.~(\ref{W-int1}),
{\it the interaction energy between dislocations} reads
\begin{align}
\label{W-int2}
W^{(AB)}&=
\int_V \bigg(
\big[\epsilon_{jkl}C_{IlMn}\epsilon_{npq}C_{RsTp}\,
F_{skMR}\big]*A^{(B)}_{Tq}\bigg)
 A^{(A)}_{Ij}\, d V\,.
\end{align}
Substituting Eq.~(\ref{phi-sol4}) into Eq.~(\ref{W-int1}),
we take an alternative formula for the interaction energy between dislocations
\begin{align}
\label{W-int3}
W^{(AB)}&=
\int_V \bigg(
C_{IlTl}\, \frac{1}{4\pi R}* A^{(B)}_{Tj}
-C_{IlTj}\, \frac{1}{4\pi R}* A^{(B)}_{Tl}\nonumber\\
&\qquad
+C_{IlMk}C_{RsTl}\, F_{skMR}* A^{(B)}_{Tj}
-C_{IlMk}C_{RsTj}\, F_{skMR}* A^{(B)}_{Tl}\bigg)
A^{(A)}_{Ij}\, d V\,.
\end{align}

\subsection{Dislocation loops}
The obtained general formulas of the previous subsections are applied to the case of dislocation loops deducing in this way the generalized Burgers equation and all other dislocation key-formulas for loops.
\par
For a dislocation loop $L$, the extended
dislocation density  and the extended plastic distortion
tensors are of the form~(see, e.g., \citep{AL13})
\begin{align}
\label{A-L}
A_{Ij}&=b_I \delta_j(L)=b_I \oint_L \delta(\Bx-\Bx')\, d L'_j\,,\\
\label{B-P}
B^{\TP}_{Ij}&=-b_I \delta_j(S)=-b_I \int_S \delta(\Bx-\Bx')\, d S'_j\,,
\end{align}
where $b_I$ is the Burgers vector in the hyperspace, $d L'_j$ denotes
the dislocation line element at
$\Bx'$ and $d S'_j$ is the dislocation loop area.
The surface $S$ is the ``cap'' over the dislocation line $L$.
The surface $S$ represents the area swept by the loop $L$ during its
motion and may be called {\it the dislocation surface}.
The plastic distortion caused by a dislocation loop is concentrated
at the surface $S$.
Thus, the surface $S$ is what determines the history of the plastic distortion
of a dislocation loop (see, e.g,~\citep{deWit60,deWit2}).
Here, $\delta_j(L)$ is the Dirac delta function for a closed curve $L$ and
$\delta_j(S)$ is the Dirac delta function for a surface $S$
whose boundary is $L$. For the forthcoming calculations we need the following relations~\citep{deWit2}
\begin{align}
\label{S-rel}
\int_V\delta_k(S')\, f(\Bx-\Bx')\, d V'=\int_S f(\Bx-\Bx')\, d S'_k\,,\\
\label{L-rel}
\int_V\delta_k(L')\, f(\Bx-\Bx')\, d V'
=\oint_L f(\Bx-\Bx')\, d L'_k\,.
\end{align}
\subsubsection{Generalized Burgers equation}
 Here, we find the expression of the extended displacement vector~(\ref{U-final}) for a dislocation loop, providing in this way the generalized Burgers equation.
\par
We start with the calculation of the first term of Eq.~(\ref{U-final}), which using the expression of $B^{\TP}_{Ij}$ (Eq.~(\ref{B-P})) and the relation~(\ref{S-rel}) becomes
\begin{align}
B^\TP_{Im,m}*\frac{1}{4\pi R}
&=-\frac{b_I}{4\pi}\,\pd_m \int_V \frac{1}{R} \, \delta_{m}(S')\, d V'
=-\frac{b_I}{4\pi} \int_S\pd_m \,\Big(\frac{1}{R}\Big) \, d S'_m
=\frac{b_I\,\Omega}{4\pi}\,,
\end{align}
where {\it the solid angle} $\Omega$ is defined by
\begin{align}
\label{Omega}
\Omega=-\int_S\pd_m \Big(\frac{1}{R}\Big)\, d S'_m
=  \int_S \frac{R_m}{R^3}\, d S'_m\,.
\end{align}
The solid angle is the angle under which the loop $L$ can be seen from
the point $\Bx$. It is very useful for the numerical implementation that the solid angle~(\ref{Omega}) can be transformed into a line integral and a constant contribution~\citep{Lazar13}
\begin{align}
\label{Omega-0-L}
\Omega=\oint_L A_k(\BR)\, d L'_k
-4\pi
\left\{
\begin{array}{rl}
\displaystyle{1}\,,\qquad &\text{if $C$ crosses $S$ positively,}\\
\displaystyle{0}\,,\qquad &\text{if $C$ does not cross $S$,}\\
\displaystyle{-1}\,,\qquad &\text{if $C$ crosses $S$ negatively.}\\
\end{array}
\right.
\end{align}
In Eq.~(\ref{Omega-0-L}), $C$ is a curve, called the ``Dirac string'',
starting at $-\infty$ and ending at the origin
(for convenience $C$ is usually chosen to be a straight line)
and $A_k(\BR)$ is the vector potential of a ``magnetic
monopole''~\citep{Eshelby, deWit81}, which is given by
\begin{align}
\label{Ak}
A_k(\BR)=\epsilon_{klm}\,\frac{\hat{n}_l R_m}{R(R+R_i \hat{n}_i)}\,,
\end{align}
where $\hat{n}_i$ is an arbitrary but constant unit vector on $C$.
\par
So far, Eq.~(\ref{U-final}) reads for a dislocation loop
\begin{align}
\label{U-ani2}
U_I(\Bx)= -\frac{b_I\,\Omega}{4\pi}-\epsilon_{rmn} C_{JkLn} F_{mkIJ}* A_{Lr}\, .
\end{align}
If we substitute Eqs.~(\ref{A-L}) and (\ref{F-int2}) into Eq.~(\ref{U-ani2})
and use the relation~(\ref{L-rel}), the final result is given by
\begin{align}
\label{U-Burgers}
U_I(\Bx)= -\frac{b_I\,\Omega}{4\pi}
+\frac{b_L \epsilon_{rmp}}{8\pi^2} \oint_L\frac{1}{R}\,  \bigg(\int_0^{2\pi}
 C_{JkLp} n_m n_k (n C n)^{-1}_{IJ}\,
d \phi\bigg)\, d L'_r
\, .
\end{align}
Eq.~(\ref{U-Burgers}) is {\it the generalized Burgers formula for dislocation loops in quasicrystals}.
It is obvious that it is the generalization of the Burgers formula
of general anisotropic elasticity (see, e.g., \citep{LK13})
towards the theory of quasicrystals. The extended displacement vector is written as the sum of a line integral plus a purely geometric part. Due to the anisotropy, an integration in $\phi$ appears in Eq.~(\ref{U-Burgers}).
In the integrand the unit vector $n_m$, which is perpendicular to $\BR$,
the expression $(n C n )_{IJ}^{-1}$ and
$C_{JkLp} n_k$ are functions of $\phi$.
It is worth noting that there is no need for further numerical differentiation.
Due to the simplicity of the result~(\ref{U-Burgers}),
it can be used directly in numerical simulations and discrete dislocation dynamics
of quasicrystals.

\subsubsection{Other dislocation key-formulas for loops}

In this subsection, we derive the generalizations of Mura-Willis equation, Peach-Koehler stress formula, Volterra equation, Peach-Koehler force and stress function tensor for a dislocation loop towards the theory of quasicrystals. Moreover, the explicit formula of the interaction energy between two dislocation loops in quasicrystalline materials is also deduced.
\par
If we substitute Eq.~(\ref{A-L}) into Eq.~(\ref{B-sol})
and perform the convolution using Eq.~(\ref{L-rel}), we
find {\it the generalized Mura-Willis equation for a dislocation loop in quasicrystals}
\begin{align}
\label{B-mura-L}
B_{Im}(\Bx)=\oint_L \epsilon_{mnr}C_{JkLn} b_L G_{IJ,k}(\BR)\, d L'_{r}\,.
\end{align}
\par
In addition, by inserting Eq.~(\ref{A-L}) into Eq.~(\ref{T-sol})
and calculate the convolution by the help of Eq.~(\ref{L-rel}),
the extended stress tensor of a dislocation loop follows
\begin{align}
\label{T-PK-L}
\Sigma_{Ps}(\Bx)=C_{PsIm}\oint_L \epsilon_{mnr}C_{JkLn} b_L G_{IJ,k}(\BR)\, d L'_{r}\,,
\end{align}
which is {\it the generalization of the Peach-Koehler stress formula} towards
the theory of quasicrystals.
\par
 Once the extended distortion tensor $B_{Im}$ (Eq.~(\ref{B-mura-L})) and the extended stress tensor $\Sigma_{Ps}$ (Eq.~(\ref{T-PK-L})) are known for a dislocation loop,  then {\it the Eshelby stress tensor for a dislocation loop} can be easily calculated by substituting Eqs.~(\ref{B-mura-L}) and  (\ref{T-PK-L}) into Eq.~(\ref{P}).
\par
By substituting the extended plastic distortion tensor~(\ref{B-P}) into
Eq.~(\ref{U-sol}) and using Eq.~(\ref{S-rel}) in order to calculate
the convolution, we obtain {\it the generalized Volterra equation for a dislocation loop in
quasicrystals}
\begin{align}
\label{U-V-L}
U_{I}(\Bx)=\int_S C_{JkLm} b_L G_{IJ,k}(\BR)\, d S'_{m}\,.
\end{align}
\par
Eqs.~(\ref{J}) and (\ref{PK}) using the relations~(\ref{A-L})
and (\ref{L-rel}) give the $J$-integral for a dislocation loop
which is equivalent to
{\it the Peach-Koehler force
for a dislocation loop $L$ interacting with the stress field $\Sigma_{Ik}$}
\begin{align}
\label{PK-L}
J_j\equiv F^{\text{PK}}_j=\int_V f^{\text{PK}}_j\, d V
=\oint_L\epsilon_{jkl} b_I \Sigma_{Ik}\, d L_l' \,.
\end{align}

Next, we consider two dislocation loops ${L^{(A)}}$ and ${L^{(B)}}$
 with Burgers vectors $b_I^{(A)}$ and $b_I^{(B)}$, respectively. If we substitute Eq.~(\ref{T-PK-L}) into Eq.~(\ref{PK-L}), {\it the Peach-Koehler force
between the dislocation loop $L^{(A)}$ in the stress field caused by the
dislocation loop $L^{(B)}$}  reads
\begin{align}
\label{PK-L-int}
F^{\text{PK}}_j=
\oint_{L^{(A)}}\oint_{L^{(B)}}
b_I^{(A)}b_T^{(B)}
\big[\epsilon_{jkl}C_{IkMn}\epsilon_{npq}C_{RsTp}\,
G_{MR,s}(\BR)\big] \, d L^{(B)}_q d L^{(A)}_l\,,
\end{align}
where $R=|\Bx^{(A)}-\Bx^{(B)}|$ is the distance between two points of
the dislocation loops  $L^{(A)}$ and  $L^{(B)}$.
\par
In Eqs.~(\ref{B-mura-L})--(\ref{U-V-L}) and Eq.~(\ref{PK-L-int}),
the gradient of the Green tensor is given by  (see Appendix~B)
\begin{align}
\label{GT-grad}
G_{IJ,k}(\BR)=
-\frac{1}{8\pi^2 R^2}\, \int_0^{2\pi}\Big(\tau_k (n C n)_{IJ}^{-1}
-n_k (n C n)_{IM}^{-1}\big[(n C \tau)_{MN}+(\tau C n)_{MN}\big]
(n C n)_{NJ}^{-1}\Big)\, d \phi\, ,
\end{align}
where $\Btau=\BR/R$ is the unit vector along $\BR$.
\par
Using Eqs.~(\ref{F-int2}) and (\ref{A-L}), Eq.~(\ref{phi-sol3}) becomes
\begin{align}
\label{phi-int}
\Phi_{Ij}(\Bx)&=\epsilon_{jkl}C_{IkMn}\epsilon_{npq}C_{RsTp}\,
\frac{b_T}{8\pi^2}\oint_L\frac{1}{R} \,
\bigg(\int_0^{2\pi} n_l n_s (n C n)_{MR}^{-1}\, d \phi\,\bigg)\, d L'_q \,.
\end{align}
Eq.~(\ref{phi-int}) gives {\it the stress function tensor for a dislocation loop in quasicrystals}. The integration in $\phi$ is to be done around $\BR$.
\par
Using Eqs.~(\ref{A-L}) and ~(\ref{L-rel}), the interaction energy~(\ref{W-int1}) becomes
\begin{align}
\label{W-int-L}
W^{(AB)}
=\oint_{L^{(A)}} b^{(A)}_I \Phi^{(B)}_{Ij} \, d {L}_j^{(A)}\,.
\end{align}
Substituting Eq.~(\ref{phi-int}) into Eq.~(\ref{W-int-L}), we obtain
{\it the interaction energy between two dislocation loops in quasicrystals}
\begin{align}
\label{W-int-L2}
W^{(AB)}
=\frac{b^{(A)}_I  b^{(B)}_T}{8\pi^2}
\oint_{L^{(A)}}\oint_{L^{(B)}}
\epsilon_{jkl}C_{IkMn}\epsilon_{npq}C_{RsTp}\,
\frac{1}{R} \,
\bigg(\int_0^{2\pi} n_l n_s (n C n)_{MR}^{-1}\, 
d \phi\,\bigg)\, d {L}^{(B)}_q\,  d {L}^{(A)}_j\,.
\end{align}
Moreover, Eq.~(\ref{W-int-L2}) may be re-written as follows
\begin{align}
\label{WAB2}
W^{(AB)}
=b^{(A)}_I b^{(B)}_T M^{(AB)}_{IT}\,,
\end{align}
where
\begin{align}
\label{MAB1-disl}
M^{(AB)}_{IT}
=\frac{1}{8\pi^2}
\oint_{L^{(A)}}\oint_{L^{(B)}}
\epsilon_{jkl}C_{IkMn}\epsilon_{npq}C_{RsTp}\,
\frac{1}{R} \,
\bigg(\int_0^{2\pi} n_l n_s (n C n)_{MR}^{-1}\, 
d \phi\,\bigg)\, d {L}^{(B)}_q\,  d {L}^{(A)}_j\,,
\end{align}
is the so-called ``dislocation mutual inductance'' tensor
(see~\citep{Kroener58,deWit60,Lardner}),
which is in general asymmetric.
\par
Furthermore, substituting Eq.~(\ref{A-L}) into Eq.~(\ref{W-int3}), we obtain an explicit formula for the interaction energy between two loops, which lends itself readily to numerical implementation
\begin{align}
\label{W-int-L3}
&W^{(AB)}
=b^{(A)}_I  b^{(B)}_T
\bigg(
C_{IlTl}
\oint_{L^{(A)}}\oint_{L^{(B)}}
\frac{1}{4\pi R}\, d {L}^{(B)}_j\,  d {L}^{(A)}_j
-C_{IlTj}
\oint_{L^{(A)}}\oint_{L^{(B)}}
\frac{1}{4\pi R}\, d {L}^{(B)}_l\,  d {L}^{(A)}_j\nonumber\\
&\
+C_{IlMk}C_{RsTl}
\oint_{L^{(A)}}\oint_{L^{(B)}} F_{skMR}(R)\, d {L}^{(B)}_j\,  d {L}^{(A)}_j
-C_{IlMk}C_{RsTj}
\oint_{L^{(A)}}\oint_{L^{(B)}} F_{skMR}(R)\, d {L}^{(B)}_l\,  d {L}^{(A)}_j
\bigg)\,.
\end{align}
\par
All the aforementioned dislocation key-formulas using Eq.~(\ref{GT-grad}) can be implemented
into numerical codes in a straightforward manner, since the appearing integrals are
``well-behaved'' functions (e.g.,~\citep{Barnett}).

\section{Conclusion}
In this work, fundamental aspects of generalized elasticity and dislocation theory of quasicrystals have been
investigated.
Generalized elasticity theory of quasicrystals is a theory of
coupled phonon and phason fields.
First,
we have pointed out the calculation of the three-dimensional elastic Green tensor for one-, two-, and three-dimensional
quasicrystals.
Second, using the Green tensor, all the dislocation key-formulas known from anisotropic elasticity, that is
Burgers formula, Mura-Willis formula, Volterra formula,
Peach-Koehler stress formula, Peach-Koehler
force, Eshelby stress tensor, the $J$-integral and the
interaction energy have been generalized towards the theory of quasicrystals.
The obtained key-formulas are important for dislocation-based plasticity since a dislocation is the elementary carrier of plasticity,
and for dislocation based fracture mechanics of
quasicrystals. Moreover, another advantage of the obtained results is that they can be used
 to build a discrete dislocation dynamics of quasicrystals
similar to the discrete dislocation dynamics of anisotropic crystals (see, e.g.,~\citep{Ghoniem}).

\section*{Acknowledgements}
The authors gratefully acknowledge grants from the
Deutsche Forschungsgemeinschaft
(Grant Nos. La1974/2-1, La1974/2-2, La1974/3-1).

\begin{appendix}
\section{The elastic Green tensor}
\label{appendixA}
\setcounter{equation}{0}
\renewcommand{\theequation}{\thesection.\arabic{equation}}

In this Appendix, we give some details concerning the calculation of the
three-dimensional Green tensor (\ref{GT})
using the inverse Fourier transform
\begin{align}
\label{G-FT-A}
G_{KM}(\BR)=\frac{1}{(2\pi)^3}\int_{-\infty}^\infty
\frac{1}{k^{2}}\, (\kappa C\kappa)^{-1}_{KM}\,
\e^{\ii \Bk\cdot\BR}\,d \Bk\,.
\end{align}
For the calculation of Eq.~(\ref{G-FT-A}), we use 
\begin{align}
\label{dk}
d \Bk=k^2 \sin\theta\, d k\, d \theta\, d \phi
\end{align}
with
\begin{align}
\Bk \cdot \BR= k\, \Bkappa\cdot\BR\,,
\end{align}
where $\Bkappa=\Bk/k$ is a unit vector, $k=|\Bk|$ and $\Bkappa=\Bkappa(\theta,\phi)$.
Considering only the real part of the integral~(\ref{G-FT-A}) since $G_{KM}(\BR)$
is a real-valued tensor function, then Eq.~(\ref{G-FT-A}) reads
\begin{align}
\label{G-FT-A2}
G_{KM}(\BR)=\frac{1}{(2\pi)^3}\int_{0}^{2\pi}
\int_{0}^{\pi}\int_{0}^{\infty}
(\kappa C\kappa)^{-1}_{KM}\,
\cos(k \Bkappa \cdot\BR)\,
d k\, \sin\theta\, d \theta\, d \phi\,.
\end{align}

On the other hand, we can first perform the $k$-integration as follows~\citep{Barton}
\begin{align}
\label{FT-k}
\int_{0}^{\infty}\cos(k \Bkappa \cdot\BR)\,
d k
=\frac{1}{2}\int_{-\infty}^{+\infty}
\e^{\ii k\, \Bkappa \cdot\BR}\,
d k
=\pi\, \delta(\Bkappa \cdot\BR)\,.
\end{align}
Since
\begin{align}
\Bkappa \cdot\BR= R \cos \theta\,,
\end{align}
we obtain by means of the property of the $\delta$-function~\citep{Barton},
$\delta(a x)=\delta(x)/|a|$,
\begin{align}
\delta(\Bkappa \cdot\BR)=\frac{1}{R} \, \delta(\cos\theta)\,.
\end{align}
In addition, we can write~\citep{Jones}
\begin{align}
\delta(\cos\theta)=
\delta(\theta-\pi/2)\,, \qquad 0< \theta <\pi\,.
\end{align}
After the $\theta$-integration, we find
\begin{align}
\label{GT-app}
G_{KM}(\BR)=
\frac{1}{8\pi^2 R}\, \int_0^{2\pi} (n C n)_{KM}^{-1}\, d \phi\, ,
\end{align}
where the integrand in Eq.~(\ref{GT-app}) must be calculated
for $\theta=\pi/2$ and $\Bkappa$ becomes $\Bn=\Bn(\phi)=\Bkappa(\pi/2,\phi)$
with $\Bn \cdot \BR=0$.
For arbitrary orthonormal vectors
$\Ba$ and $\Bb$ in the plane $\Bn \cdot \BR=0$, the vector $\Bn$ can be given
by (see~\citep{Teodosiu,Barnett})
\begin{align}
\Bn=\Ba \cos\phi+\Bb \sin\phi\,.
\end{align}

\section{The first derivative of the Green tensor}
\label{appendixB}
\setcounter{equation}{0}
\renewcommand{\theequation}{\thesection.\arabic{equation}}

In this Appendix, we calculate the  first derivative of the Green tensor~(\ref{GT})
\begin{align}
\label{GT-A}
G_{IJ}(\BR)=
\frac{1}{8\pi^2 R}\, \int_0^{2\pi} (n C n)_{IJ}^{-1}\, d \phi\, .
\end{align}
The differentiation of Eq.~(\ref{GT-A}) with respect to $x_k$ gives
\begin{align}
\label{GT-A-2}
G_{IJ,k}(\BR)=
\frac{1}{8\pi^2}\, \int_0^{2\pi}
\bigg((n C n)_{IJ}^{-1}\, \Big(\frac{1}{R}\Big){,_k}
+\frac{1}{R}\, (n C n)_{IJ,k}^{-1}\bigg)\, d \phi\, .
\end{align}
Using the relations~\citep{Balluffi}
\begin{align}
\label{Rel1}
 (n C n)_{IJ,k}^{-1}=-(n C n)_{IM}^{-1}\, (n C n)_{MN,k}\,  (n C n)_{NJ}^{-1}\,,
\end{align}
\begin{align}
\label{Rel2}
\Big(\frac{1}{R}\Big){,_k}=-\frac{R_k}{R^3}=-\frac{\tau_k}{R^2}\,,
\end{align}
\begin{align}
\label{Rel3}
(n C n)_{MN,k}=-\frac{n_k}{R}\, \big[(n C \tau)_{MN}+(\tau C n)_{MN}\big]\,,
\end{align}
\begin{align}
\label{Rel4}
\delta n_k =-\frac{1}{R}\, n_j\delta x_j\, \tau_k\,,
\end{align}
Eq.~(\ref{GT-A-2}) becomes
\begin{align}
\label{GT-grad-A}
G_{IJ,k}(\BR)=
-\frac{1}{8\pi^2 R^2}\, \int_0^{2\pi}\Big(\tau_k (n C n)_{IJ}^{-1}
-n_k (n C n)_{IM}^{-1}\big[(n C \tau)_{MN}+(\tau C n)_{MN}\big]
(n C n)_{NJ}^{-1}\Big)\, d \phi\, ,
\end{align}
where $\Btau=\BR/R$.
Eq.~(\ref{GT-grad-A}) is analogous to the corresponding expression in
anisotropic elasticity (see, e.g.,~\citep{Lothe,Balluffi}).

\end{appendix}

\end{document}